\documentstyle[preprint,prl,aps]{revtex} % for preprint form
\begin{document}
\draft
% \twocolumn
\title{Coupling Between Planes and Chains in
$\bf YBa_{2}Cu_{3}O_{7}$ : A Possible Solution for the
Order Parameter Controversy}
\author{R.\ Combescot and X.\ Leyronas}
\address{Laboratoire de Physique Statistique, Ecole Normale Sup\'erieure,
24 rue Lhomond, 75231 Paris Cedex 05, France}
\date{Received \today}
\maketitle
\begin{abstract}
We propose to explain the contradictory experimental evidence
about the symmetry of the order parameter in $YBa_{2}Cu_{3}O_{7} $
by taking into account the coupling between planes and chains.
This leads to an anticrossing of the plane and chain band. We
include an attractive pairing interaction within the planes
and a repulsive one between planes and chains, leading to opposite
signs for the order parameter on planes and chains, and to nodes
of the gap because of the anticrossing. Our model blends s-wave
and d-wave features, and provides a natural explanation for
all the contradictory experiments
\end{abstract}
\narrowtext
The mechanism responsible for pairing in high $T_{c } $ superconductors
is presently the subject of a vivid debate \cite{[1]}. There is a high
suspicion that some, if not all, high $T_{c } $ compounds are unconventional.
On one hand a repulsive interaction between electrons can lead
to the formation of pairs, as in the case where the interaction
is due to the exchange of spin fluctuations \cite{[2]}. For this specific
case the order parameter has a d-wave symmetry which implies
in particular that it changes sign on the Fermi surface. On
the other hand an attractive interaction between electrons leads
to an  order parameter which does not change sign on the Fermi
surface, even if there is a sizeable anisotropy of the gap.
In this case the order parameter has an s-wave symmetry as for
the conventional mechanism of phonon mediated pairing. Therefore
an efficient way to check the pairing mechanism, and to decide
if high $T_{c } $ superconductors are conventional or unconventional,
is to look experimentally at the symmetry of the order parameter.
  \par
  \bigskip
Very surprisingly, recent experiments performed in $YBa_{2}Cu_{3}O_{7} $
in order to settle this issue of the order parameter symmetry
have given convincing, but contradictory, answers \cite{[1]}. For example
the observation of a sizeable Josephson current \cite{[3]} in a c-axis
tunnelling junction between $YBa_{2}Cu_{3}O_{7} $ and  Pb is
quite difficult to reconcile with a pure d-wave symmetry. Similarly
the absence of angular dependence and of sign change in the
critical current of YBCO - $   YBCO $ grain boundary junctions
in the a-b plane \cite{[4]} $   is $ clearly $   in $ favor of an s-wave
interpretation. On the other hand the evidence for a $\pi  $ shift
in corner SQUID experiments \cite{[5]} is a clear indication for a
change of sign of the order parameter between the a and the
b axis. Similarly the observation of a spontaneous magnetization
corresponding to a half magnetic flux quantum in 3 grain-boundary
Josephson junctions \cite{[6]} favors clearly d-wave symmetry. Finally
all the experiments which show an evidence for low energy excitations
at low temperature, such as tunnelling \cite{[1]}, linear temperature
dependence of the penetration depth in YBCO crystals or films
\cite{[7]}, or photoemission experiments in $Bi_{2}Sr_{2}CaCu_{2}O_{8} $
 \cite{[8]} are also more naturally interpreted in terms of d-wave
symmetry, although they are consistent with a strongly anisotropic
s-wave order parameter. While there is always the possibility
that one set of experiments is spoiled for obscure reasons,
this is rather unlikely in view of the quality, the independence
and of the variety of these experiments. A more natural view
is to look for an explanation which allows to reconcile these
various results.   \par
  \bigskip
It is just our purpose in this paper to present such a model,
which blends s-wave and d-wave features, and provides a natural
explanation for all the contradictory experiments \cite{[1]}. The basic
ingredient of our model is the coupling between planes and chains.
This is taken into account both in the band structure of YBCO
(diagonal coupling), and in the (repulsive) pairing interaction
between planes and chains (off-diagonal coupling). Indeed in
order to account for all the experiments in favor of a d-wave
interpretation, we have to take an order parameter which changes
sign. As in the spin fluctuation mechanism, the natural explanation
for this is the existence of a repulsive interaction between
electrons. Repulsive interactions are actually already present
in conventional superconductivity because of the existence of
the Coulomb repulsion between electrons, which is taken into
account by the Coulomb pseudopotential. This leads indeed to
a change of sign of the order parameter, but this occurs in
the frequency dependence rather than in the wavevector dependence.
Therefore if the Coulomb interaction is the dominant mechanism
for scattering between some pieces of the Fermi surface, this
can lead to a change of sign in the order parameter on the Fermi
surface, in close analogy with the spin fluctuation mechanism.
Similar ideas have been proposed very recently, pointing out
that the sign of the order parameter could be opposite on two
different sheets of the Fermi surface, due to repulsive interaction
produced by spin fluctuation or by direct Coulomb interaction
\cite{[9]}. These two sheets could be produced by the two bands corresponding
to the two $CuO_{2} $ planes, or they could correspond to the
plane band and the chain band. Here we will take the view that,
due for example to Coulomb interaction, the order parameter
has opposite sign on the plane band and on the chain band. For
simplicity we will consider here a single plane band.   \par
  \bigskip
However, although this change of sign of the order parameter
on different sheets of the Fermi surface can account for the
experiments showing a $\pi  $ shift in Josephson junctions, it does
not explain a number of facts in favor of a d-wave like interpretation.
These are all the experiments which point toward the existence
of low lying excitations at low temperature \cite{[1],[7],[8]}, and therefore
toward the presence of nodes of the gap on the Fermi surface.
We propose to explain this feature by taking into account the
coupling between plane and chain occuring in the band structure.
This coupling is likely due to the O4 apical oxygen. Physically
this means that an electronic eigenstates is never purely a
plane electron or a chain electron. Rather it has a mixed nature
with components on both plane and chain.  This coupling is a
well known feature in band structure calculations \cite{[10]}. Naturally
this coupling is fairly small, and it is of importance only
when the plane and the chain band intersect, where it leads
to a standard anticrossing feature in the dispersion relations
found in all band structure calculations. Similarly, wherever
the (uncoupled) pieces of the Fermi surface related to plane
and chain cross, the coupling leads to an anticrossing as it
is also seen in band structure calculations \cite{[10]}. This anticrossing
has the important consequence that, when we move on a given
sheet of the Fermi surface, we go from a part which corresponds
physically to a plane electron, to a part which coresponds physically
to a chain electron. We will take the following Hamiltonian
as a model for the band structure:
\begin{eqnarray}
{\rm H}_{0}=\sum\nolimits\limits_{k}^{}
{\varepsilon }_{k}{c}_{k}^{+}{c}_{k}+\sum\nolimits\limits_{k}^{}
 {\varepsilon'}_{k}{d}_{k}^{+}{d}_{k}+\sum\nolimits\limits_{k}^{}
{T}^{}{c}_{k}^{+}{d}_{k}+h.c.
\label{eq1}
\end{eqnarray}
It describes semi-quantitatively all the features of the band
structure which are important for us.  Here $c_{k }^{+} $ and
$d_{k }^{+} $ are creation operators in the plane and in the chain
band respectively. We take $\epsilon _{k } $  = - $2t_{0} $ ( $cos(k_{x }a) $
+ $cos(k_{y }a) $ ) + $2t_{0} $ $cos(k_{x }a) $ $cos(k_{y }a) $ - $\mu _{0} $
   and $\epsilon ^{\prime}_{k } $  =  - $2t_{1} $ $cos(k_{y }a) $ - $\mu _{1} $
 , with
$t_{0} $  = 0.33 eV, $t_{1} $  = 0.53 eV, $\mu _{0} $  = - 0.46 eV and
$\mu _{1} $  = - 0.74 eV, in order to obtain a Fermi surface in
reasonable agreement with band structure calculations, but this
is not essential for our purpose. We neglect the {\bf k} dependence
of the plane-chain coupling T, since this term is relevant only
in a rather small region in {\bf k} space. Our model is purely
2-dimensional and we have neglected any 3-dimensional effect,
since they are inessential for our purpose. The energy $e_{\pm } $
 of the eigenstates of Hamiltonian Eq.(1) are 2 $e_{\pm } $ = $\epsilon  $
 + $\epsilon ^{\prime} $ $\pm  $ [ ( $\epsilon  $ - $\epsilon ^{\prime} $
$)^{2} $
+ 4 $T^{2} $ $]^{1/2} $ and the corresponding
Fermi surface e $_{\pm } $  = 0 is shown in Fig.1 for the case T
= 0.1 eV.  \par
  \bigskip
Let us now consider the superconducting state. In the planes
we take a purely attractive pairing interaction, which can be
for example of phononic origin. On the other hand the off-diagonal
coupling between plane and chain is repulsive, as indicated
above. Finally we have to consider for generality a pairing
interaction in the chains. Neglecting all unnecessary wavevector
dependence, this leads us in weak coupling to the following
interaction ( with singlet pairing understood ) :
\begin{eqnarray}
{\rm H}_{int}=-g\sum\nolimits\limits_{k,k'}^{}
{c}_{k'}^{+}{c}_{-k'}^{+}{c}_{-k}{c}_{k}+K\sum\nolimits\limits_{k,k'}^{}
{d}_{k'}^{+}{d}_{-k'}^{+}{c}_{-k}{c}_{k}+h.c.+g'\sum\nolimits\limits_{k,k'}^{}
{d}_{k'}^{+}{d}_{-k'}^{+}{d}_{-k}{d}_{k}
\label{eq2}
\end{eqnarray}
where g and K are positive. This gives finally the following
mean-field Hamiltonian :
\begin{eqnarray}
{\rm H}^{}={H}_{0}+\Delta \sum\nolimits\limits_{k}^{}
{c}_{k}^{+}{c}_{-k}^{+}+\Delta '\sum\nolimits\limits_{k}^{}
{d}_{k}^{+}{d}_{-k}^{+}+h.c.
\label{eq3}
\end{eqnarray}
where we have set:
\begin{eqnarray}
\rm \Delta =-g\sum\nolimits\limits_{k}^{}
<{c}_{-k}{c}_{k}>+K\sum\nolimits\limits_{k}^{} <{d}_{-k}{d}_{k}>
\label{eq4}
\end{eqnarray}
\begin{eqnarray}
\rm \Delta '=K\sum\nolimits\limits_{k}^{}
<{c}_{-k}{c}_{k}>+g'\sum\nolimits\limits_{k}^{} <{d}_{-k}{d}_{k}>
\label{eq5}
\end{eqnarray}
It is now easy to see what happens physically. Because the pairing
is attractive in the planes, but repulsive between plane and
chain, the order parameters $\Delta  $ and $\Delta ^{\prime} $
( which correspond physically
to the order parameter in the uncoupled planes and chains respectively
) will have opposite sign. On the other hand, since ( because
of the anticrossing ) when we move on a single sheet of the
Fermi surface, we go continuously from a plane like part to
a chain like part  we see ( Fig.1 ) that on moving on this given
sheet the order parameter will necessarily change sign, since
it will be essentially equal to $\Delta  $ on the plane like part, and
to $\Delta ^{\prime} $ on the chain like part. This implies by continuity that
the order parameter has necessarily a node on each sheet of
the Fermi surface.  \par
  \bigskip
 This is easily shown explicitely by solving exactly for the
excitation spectrum of the Hamiltonian Eq.(3). One finds two
branches $E_{\pm }(${\bf  k}) for this spectrum :
\begin{eqnarray}
\rm 2{E}_{\pm }^{2}=\eta +\eta '\pm
\sqrt {(\eta -\eta '{)}^{2}+4({\tau}^{2}+{\delta }^{2})}
\label{eq6}
\end{eqnarray}
where $\eta  $ = $\epsilon ^{2}+ $ $\Delta ^{2}+ $ $T^{2} $  and $\eta
^{\prime} $
= $\epsilon ^{\prime   2}+ $ $\Delta ^{\prime   2}+ $
$T^{2}, $  $\delta  $ = T ( $\Delta  $ - $\Delta ^{\prime} $ ) and $\tau  $ = T
(
$\epsilon  $ + $\epsilon ^{\prime} $ );  $\epsilon  $ and
$\epsilon ^{\prime} $ are for $\epsilon _{k } $    and $\epsilon ^{\prime}_{k }
$
 . One obtains indeed that
 E $_{-} $  ({\bf k})  =  0 if  $\epsilon  $ $|\Delta ^{\prime}| $  =
$\epsilon ^{\prime} $
$|\Delta \vert  $  = $\pm  $ [
$|\Delta  $ $\Delta ^{\prime}| $ ( $T^{2} $ - $|\Delta  $ $\Delta ^{\prime}|) $
$]^{1/2} $
 and  $\Delta  $ $\Delta ^{\prime} $ $< $ 0. The
wavevectors of these zero energy excitations are actually slightly
off the normal state Fermi surface, which is given by  $\epsilon  $  $\epsilon
^{\prime}= $
$T^{2} $  . But the departure is small when the band coupling
is large compared to the order parameter  $T^{2} $ $>> $  $|\Delta  $ $\Delta
^{\prime}|. $
 On the other hand it is very interesting to note that these
zero energy excitations fuse at  $\epsilon  $ =  $\epsilon ^{\prime} $  = 0 for
$T^{2} $
=
$|\Delta  $ $\Delta ^{\prime}| $ , and they disappear for $T^{2} $ $< $
$|\Delta  $
$\Delta ^{\prime}| $ . However
in this regime $T^{2} $ $< $ $|\Delta  $ $\Delta ^{\prime}|, $ although there
are
no longer
nodes in the gap, the coupling will produce a depression of
the gaps of each band in the anticrossing regions. Naturally
one recovers for T = 0 the excitation spectrum
 $E^{2}_{\pm }(${\bf  k}) = $\epsilon ^{2}+ $ $\Delta ^{2} $  and  $\epsilon
^{\prime   2}+ $
$\Delta ^{\prime   2} $
 for uncoupled bands. In the regime $T^{2} $ $>> $  $|\Delta  $ $\Delta
^{\prime}| $
, it
is easier to first diagonalize the normal state Hamiltonian
which leads to eigenstates with energies $e_{\pm } $ . One takes
then into account the pairing Hamiltonian, neglecting the coupling
it induces between the bands $e_{\pm }. $ This leads naturally to
the introduction of a {\bf k} dependent order parameter $\Delta _{k } $
in each band.  We find in this limit  $\Delta _{k } $ = ( $\Delta  $ $\epsilon
^{\prime} $
+
$\Delta ^{\prime} $ $\epsilon  $ )/( $\epsilon  $ + $\epsilon ^{\prime}), $
valid
on the two sheets of the Fermi surface,
corresponding to each of the two bands. Since on a given sheet
of the Fermi surface one goes from $\epsilon  $ $>> $ $\epsilon ^{\prime} $ to
$\epsilon  $
$<< $ $\epsilon ^{\prime} $ ,
one sees explicitely that the order parameter goes continuously from
$\Delta  $  to $\Delta ^{\prime} $ or vice-versa , and therefore
it changes sign for $\Delta  $
$\Delta ^{\prime} $ $< $ 0. Naturally the values of  $\Delta  $ and
$\Delta ^{\prime} $  are  obtained
from the couplings g, g' and K by the gap equations Eq.(4) and
(5). On Fig.2 we show the result for the variation of the order
parameter for the arbitrary choice $\Delta  $ = 1 and $\Delta ^{\prime}
$ = - 0.5 (arbitrary  units ).
The corresponding positions of the nodes of the gap
are indicated in Fig.1. We will now consider the physical implications
of our model and show that it solves a number of problems and
accounts for many experiments.  \par
  \bigskip
Our model mixes features from s-wave physics and from d-wave
physics, and it is therefore quite unconventional. It is d-wave
like because there are repulsive interactions between different
parts of a given sheet of the Fermi surface, leading to a change
of sign of the order parameter and nodes in the gap. The global
result mimics closely d-wave pairing. However in contrast with
d-wave pairing, the nodes do not satisfy $k_{x } $ = $\pm  $ $k_{y }, $
since they are rather in the anticrossing regions ( which happen
to be not so far from $k_{x } $ = $\pm  $ $k_{y } $  in YBCO ). Because
of the two sheets of the Fermi surface we have 8 nodes instead
of 4. On the other hand our model is s-wave like because the
plane pairing is purely attractive. But in contrast to standard
superconductivity the repulsive interaction contributes to raise
$T_{c } $  instead of destroying it. Finally we obtain a strong
anisotropy in the gap, although we do not need to invoke any
anisotropy in the planes.  \par
  \bigskip
The c-axis Josephson experiment \cite{[3]} is easily accounted for
by our model, since the strengths $|\Delta \vert  $ and $|\Delta ^{\prime}\vert
 $
 of the order
parameter with different signs are not related by any symmetry.
Therefore there is no reason for complete cancellation of the
total current between different parts of the Fermi surface,
while a partial cancellation may explain why the observed current
is only \~15% of the expected value. On the other hand, because
the chain electrons have their velocity essentially along the
b-axis, they can naturally give the dominant contribution to
tunnelling in this direction, while only plane electrons ( with
opposite sign for the order parameter ) will contribute to tunnelling
along the a-axis. Therefore the situation is in this respect
the same as with d-wave pairing, and our model accounts naturally
for the $\pi  $ shift in corner SQUID experiments \cite{[5]}, as well as
for the spontaneous magnetization $\Phi _{0}/2 $  found in 3-grain
boundary junctions \cite{[6]}. We provide also a possible explanation
for the s-wave result of the hexagonal Josephson experiment
\cite{[4]}. Indeed because of the specific choice of geometry, there
is the possibility that no boundary is perpendicular to the
b-axis. This may imply that, because the chain electrons have
their velocity along b, their contribution is strongly reduced
(or at least overwhelmed by opposite sign contributions, in
case of twinning for example ) leaving only the s-wave like
contribution from the plane electrons. Finally because of the
nodes in the gap our model explains naturally all the experiments
in favor of low energy excitations (''states in the gap'') such
as the linear temperature dependence of the penetration depth,
tunnelling experiments, Raman scattering,NMR \cite{[1]}.  \par
  \bigskip
The chain band is essential in our model in order to explain
the peculiar properties of YBCO. It is clear that, if there
is some sizeable disorder in the chains, as produced for example
by loss of 01 oxygen atoms, the anticrossing which occurs in
specific regions of {\bf k} space will be strongly perturbed
or destroyed. We can model this effect by a decrease of the
coupling T between plane and chain bands, since indeed the effect
of the chain on the plane will be reduced. As we have seen,
this may lead to the disappearance of the nodes of the gap.
And indeed only good YBCO cristals display a strong anisotropy
of the penetration depth  \cite{[11]} (proving metallic behaviour of
the chains ) and a linear T dependence of the penetration depth
\cite{[7]} (proving the existence of nodes ). On the other hand, as
we have seen, even if the nodes have disappeared, a local decrease
of the gap will survive as a remnant if there is some coupling
left. In this way our model provides a simple explanation for
(strongly) anisotropic s-wave superconductivity.  This may account
for the observation of a small gap in penetration depth \cite{[12]}
or in tunnelling experiments. Our explanation for the behaviour
of $\lambda (T) $ is naturally consistent with the fact that $\lambda (T) $ is
BCS-like in $Nd_{1.85}Ce_{0.15}Cu0_{4} $ \cite{[13]} where only
$Cu0_{2} $  planes are present. Quite generally the systematic
study of the effect of desoxygenation in YBCO will be a crucial
test of validity for our model. We note in this respect that,
in $Bi_{2}Sr_{2}CaCu_{2}0_{8}, $ the Bi0 plane bands is coupled
to the CuO plane bands \cite{[14]} and that our model, where the role
of the CuO chains would be played by the BiO planes, could also
explain the strong anisotropy of the gap observed in photoemission
experiments \cite{[8]}.  \par
  \bigskip
Finally our model accounts naturally for the fact that $T_{c } $
 in YBCO is rather insensitive to impurities \cite{[1]} ( except for
magnetic ones ). Indeed, since anticrossing occurs only in restricted
regions of {\bf k} space, we expect physically that impurity
scattering is dominantly plane-plane or chain-chain, and not
plane-chain. Therefore electrons are not scattered between parts
of the Fermi surface having order parameter with opposite sign,
and therefore the physical reason for the reduction of $T_{c } $
 mostly disappears. In contrast non magnetic impurity scattering
( as well as electron-phonon interaction ) acts as a pair-breaker
in d-wave \cite{[15]}, and the insensitivity of $T_{c } $ requires more
specific explanations. We note that we have used the same physical
reason to restrict the attractive pairing interaction to the
planes.   \par
  \bigskip
Let us conclude by some remarks on the dependence of the critical
temperature $T_{c } $  on the repulsive interaction K. We have
no experimental indication on whether the pairing interaction
g' is attractive or repulsive on the chains. Therefore let us
take g' = 0 for simplicity. Since the summations over {\bf k}
in Eq.(4) and (5) are over the whole Fermi surface, whereas
anticrossing occurs only in restricted regions of {\bf k} space,
we can to a first approximation perform the calculation to zeroth
order in the band coupling T. Introducing cut-off $\omega _{\Delta } $ and
$\omega _{c } $ for the attractive and repulsive interactions g and
K, the weak coupling calculation leads to:
\begin{eqnarray}
{\rm 2 \over \ell n(1.13 {\omega }_{c}/{T}_{c})}={\lambda +\sqrt {{\lambda
}^{2}+4kk' (1+\lambda  \ell n({\omega }_{c}/{\omega }_{D}))}
\over 1+\lambda  \ell n({\omega }_{c}/{\omega }_{D})}
\label{eq7}
\end{eqnarray}
where $\lambda  $ = N g , k = N K and k' = N'K , with N and N' being
the density of states of the (uncoupled) plane band and chain
band respectively. Naturally we find that the repulsive interaction
increases the critical temperature, compared to the situation
where we would only have the attractive s-wave coupling $\lambda  $ at
work. It is tempting to compare this increase with the increase
of $T_{c } $ when the chains in YBCO are build-up upon adding oxygen.
We may speculate for example that, at the 60 K plateau, only
the attractive interaction is working, and that the increase
up to 92 K is due to the progressive increase of the repulsive
interaction caused by the chain construction, although there
is also a contribution due to the variation of ''doping''. We
can also make a rough estimate of the isotope effect from Eq.(7),
which gives  $(\delta \omega _{D }/\omega _{D }) $
 / $(\delta T_{c }/T_{c }) $ = 1 +
2 $(kk'/\lambda ) $ ln $(1.13 \omega _{c } $  / $T_{c }) $ .
As expected, the Coulomb
repulsive interaction leads to a sizeable decrease in the isotope
effect. This may explain the small oxygen isotope effect observed
in $YBa_{2}Cu_{3}O_{7}, $ while it is found to be much stronger
for lower $T_{c } $ superconductors such as $La_{2-x }Sr_{x }CuO_{4}. $
Therefore we obtain, in agreement with experiment, an increase
of $T_{c } $ upon oxygenation while the isotope effect is decreasing
at the same time. In this respect our model behaves as expected
from a mixed mechanism where both electron-phonon interaction
and Coulomb repulsion contribute to pairing.   \par
  \bigskip
We are grateful to N. Bontemps and L.A. de Vaulchier for numerous
and useful discussions.    \par

\begin{figure}
\caption{Fermi surface in the Brillouin zone. The sign $\pm  $ of the
order parameter on the various parts of the Fermi surface is
indicated. The filled circles give the positions of the nodes
of the gap.}
\label{Fig1}
\end{figure}
\begin{figure}
\caption{Variation of the order parameter $\Delta _{k } $ on the two
sheets of the Fermi surface for $k_{x } $ and $k_{y } $ $> $ 0, as a
function of the angular position $\theta  $ = atan $(k_{y } $  $/k_{x } $).}
\label{Fig2}
\end{figure}
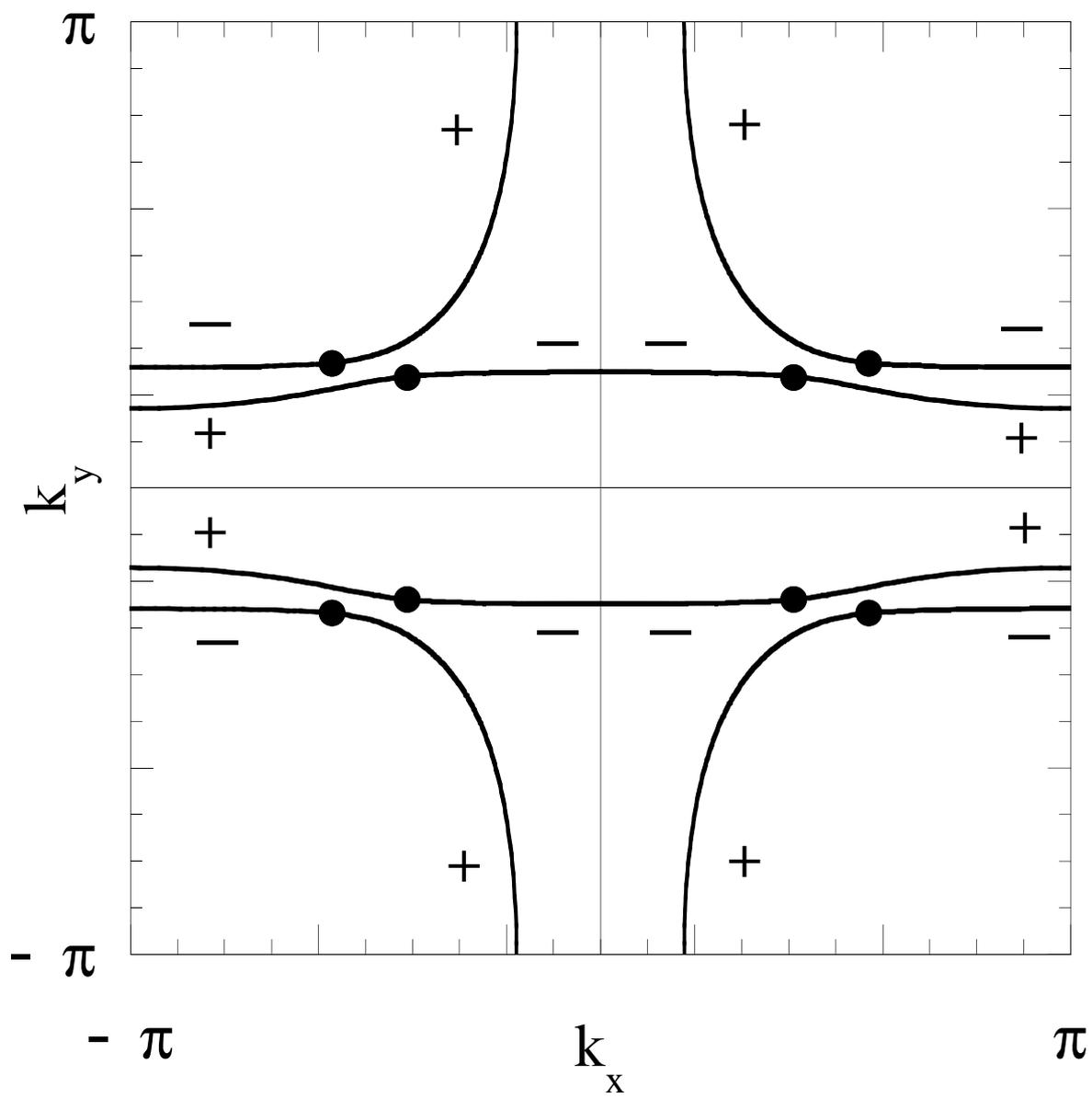
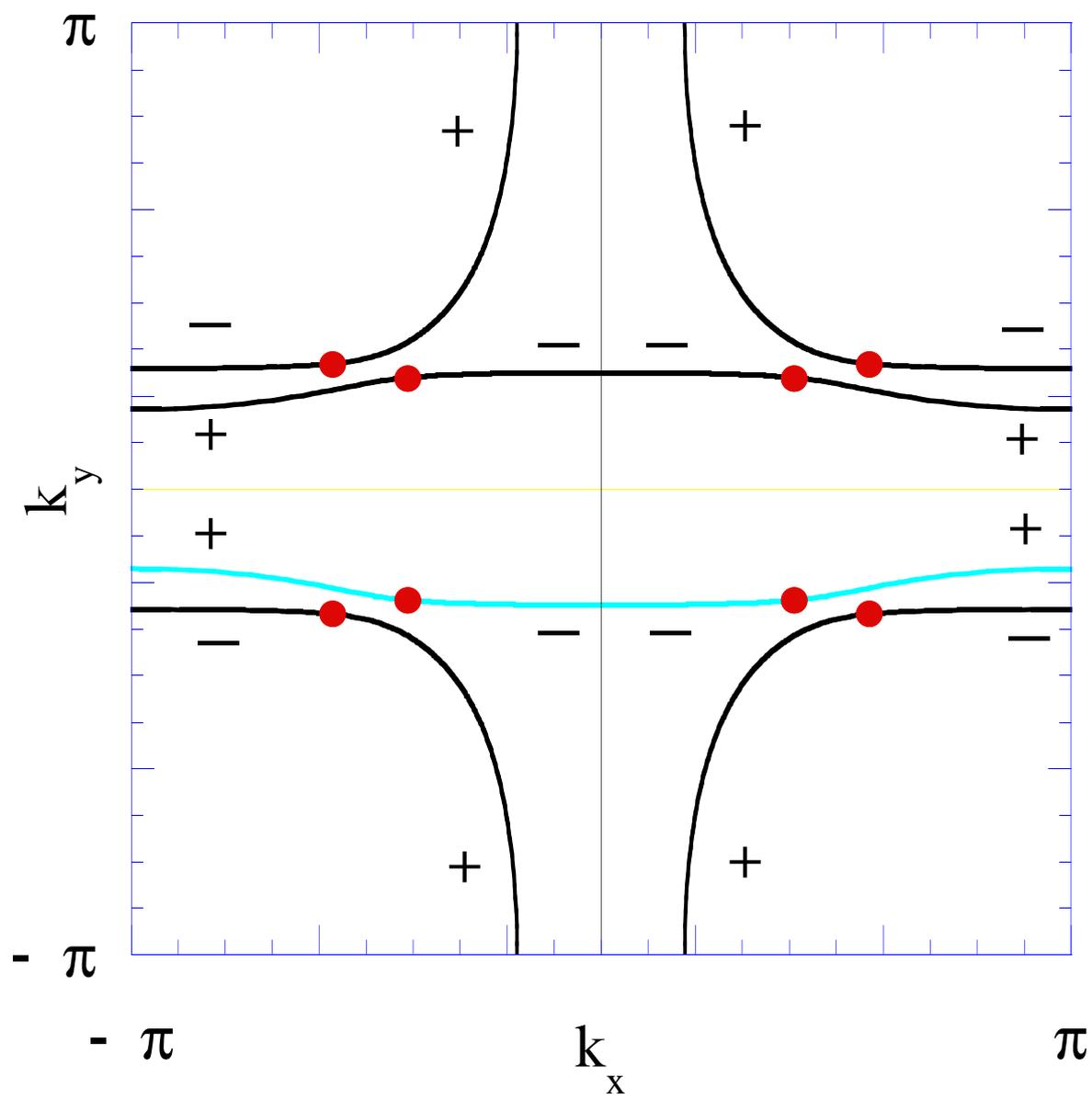
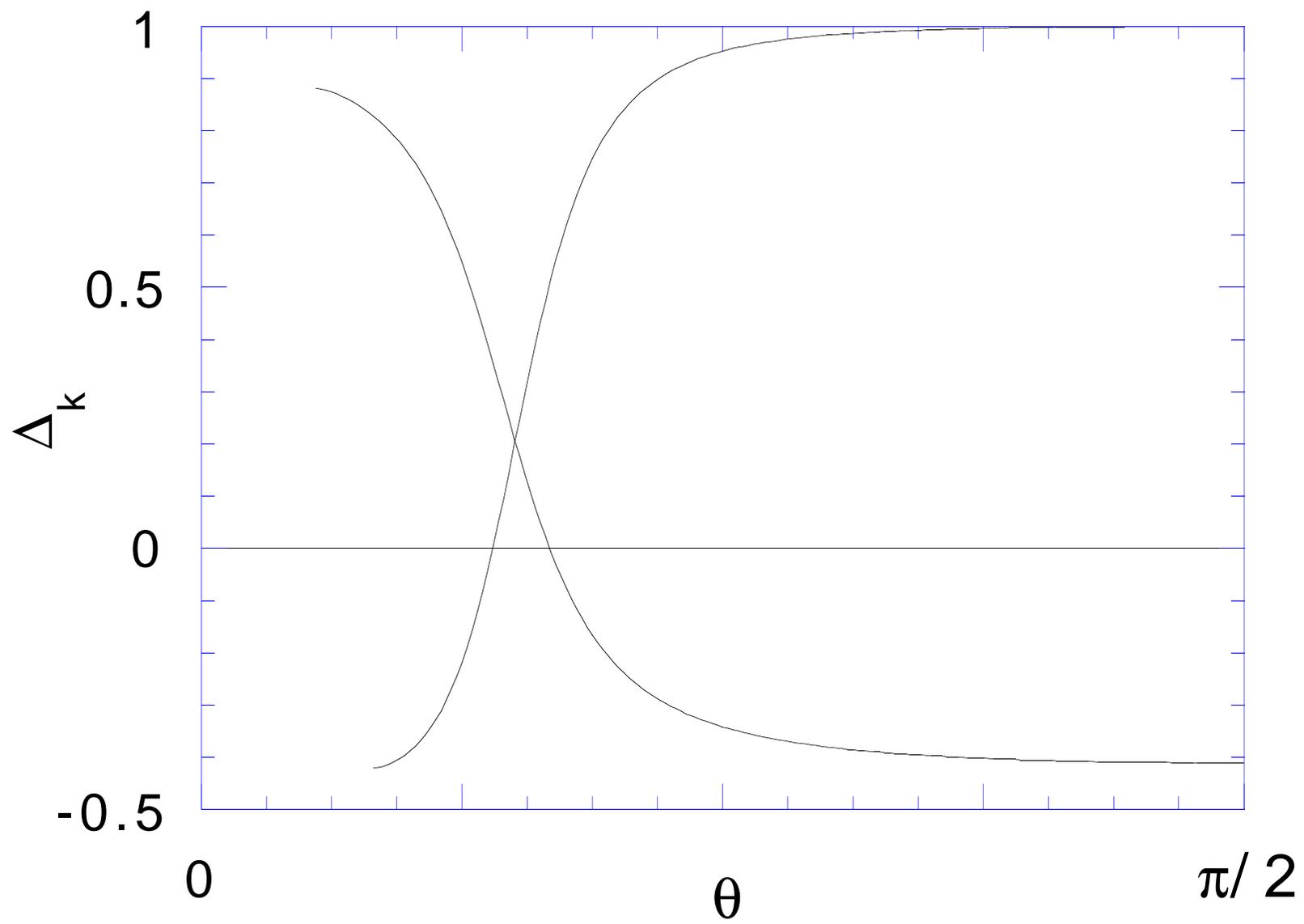
\end{document}